\documentclass[12pt,a4paper]{article}

\usepackage{cite}
\usepackage{latexsym,amssymb,amsmath,amscd,amsthm,amsxtra,graphicx}

\newcommand{\ba}{\begin{eqnarray}}
\newcommand{\ea}{\end{eqnarray}}
\newcommand{\beq}{\begin{equation}}
\newcommand{\eeq}{\end{equation}}
\newcommand{\bc}{\begin{center}}
\newcommand{\ec}{\end{center}}

\begin{document}

\title{Scalar waves from a star orbiting a BTZ black hole}
\author{Xing-Hua Wu, Ran Li, Jun-Kun Zhao\\
{\normalsize \it Department of Physics, Henan Normal University,
 Xin Xiang, 453007, China}}
\date{}
\maketitle

\begin{abstract}
In this paper we compute the decay rates of massless scalar waves 
excited by a star circularly orbiting around the non-extremal (general) and extremal
BTZ black holes. These decay rates are compared with the corresponding
quantities computed in the corresponding dual conformal field theories respectively.
We find that matches are achieved in both cases.

\end{abstract}

\section{Introduction}

In the conjecture of AdS/CFT correspondence \cite{ads-cft}, 
an issue frequently examined is how far one can go under this correspondence.
Lots of evidences of this correspondence have been achieved by matching
of corresponding quantities computed on two sides. 

For example, decay rates of various propagating
waves on a variety of spacetime backgrounds are compared to the decay rates
computed in corresponding dual CFT's 
\cite{ms_low_energy_dynamics,bhss_superradiance_kerr_cft,dd_IIB_T4_S1,moz_btz,bss_btz}.
Quasi-normal modes of black hole spacetimes are matched to the poles
of correlation functions of CFT \cite{bss_qnm}.

Recently, decay rates of excitations induced by orbiting star near the horizon
of extremal-rotating black hole have been compared with the corresponding
quantities computed in the boundary CFT \cite{ps_kerr_cft,bgx_scalar_KN}. 
Since the near-horizon limit of rotating
black holes are expected to be dual to some CFT's on the boundary, 
e.g., the Kerr/CFT correspondence, the matching of the decay rates computed on 
two sides is natural.

A general (non-extremal rotating) BTZ black hole \cite{btz_metric} 
is expected to be dual to a CFT on the boundary controlled by 
two copies of Virasoro algebras: left and right movers.
The matching of decay rates in Ref. \cite{dd_IIB_T4_S1} and the matching of
quasinormal modes to poles in Ref. \cite{bss_qnm}, among many others, are 
successful examples of the BTZ/CFT correspondence.
So an interesting question is whether some
matching can be achieved for this BTZ/CFT correspondence
if one adding to the system an orbiting source just
like what is done in \cite{ps_kerr_cft,bgx_scalar_KN}. Sect.2
will discuss this question.

Near horizon extremal BTZ black hole is expected to be dual to
a chiral 2D CFT \cite{strominger_ads2_qg,bbjs_XCFT_Ebh}. So it
is also interesting to ask, when an orbiting star is added
to the black hole system, whether the matching of decay rates computed
in the two sides can be achieved. This will be investigated in Sect.3.

For the BTZ spacetime, it is simple
enough to give analytic results. On the other hand, it appears frequently
as near horizon portion of other interesting spacetime, e.g., from various supergravity
models. So explorations of BTZ/CFT correspondence from different
viewpoints are very useful.

\section{Non-extremal BTZ/Non-chiral CFT}

\subsection{Decay rate on black hole side}

In this subsection, we will compute the decay rate of massless scalar wave
excited by a star circularly orbiting around the BTZ black hole.

The BTZ black hole is a solution to the vacuum Einstein equations
in 2+1 dimensions with a negative cosmological constant $\Lambda =
-1/\ell^2$.  In Schwarzschild-like coordinates, the BTZ metric is \cite{btz_metric}
\beq
ds^2 = -N^2dt^2 + N^{-2}dr^2
  + r^2\left( d\phi + N^\phi dt\right)^2
\label{btz-metric}
\eeq
with lapse and shift functions
\ba
N^2 &=& {(r^2-r_+^2)(r^2-r_-^2)\over \ell^2\,r^2} ,
\\[5pt]
N^\phi &=& - {r_+ r_-\over\ell\, r^2} 
\label{btz-lapse}
\ea
where the $r_+$ and $r_-$ are the locations of the outer and inner horizons
and related to the mass and the angular momentum of the black hole:
\beq
M={r_+{}^2+r_-{}^2\over8G\ell^2}, \qquad J={r_+ r_-\over4G\ell} \ .
\label{m-j}
\eeq
The Hawking temperature is
\ba
T_{\rm H}={r_+^2-r_-^2\over 2\pi r_+\ell}
\ea
For convenience in the following, one can define the following coordinate,
\cite{bss_qnm}
\ba
x^+ = {1\over\ell}(r_+ t/\ell - r_-\phi)\,,\quad
x^-={1\over\ell}(-r_- t/\ell + r_+\phi)\,,\quad
z={r^2-r_+^2\over r^2-r_-^2}
\ea
Only the dynamics outside the event horizon, $r>r_+$, will be considered in
this paper, so we have $0<z<1$.
In this coordinate the metric becomes,
\ba
ds^2= \ell^2 \left( -{z\over 1-z}(dx^+)^2 + {1\over 1-z}(dx^-)^2 
+ {1\over 4z(1-z)^2}dz^2
\right)
\ea
Now consider a star moving outside the black hole along
a circular orbit with constant radial coordinate, $z_*=z_0$. Its
conserved energy and angular momentum can be determined by two
Killing vectors, $\xi_{(t)}=\partial_t$ and $\xi_{(\phi)}=\partial_\phi$,
\ba
E = -g_{\mu\nu}u^\mu\xi_{(t)}^\nu\,,\qquad
L = g_{\mu\nu}u^\mu\xi_{(\phi)}^\nu
\ea
where $u^\mu=\dot x_*^\mu={d x_*^\mu\over d\tau}$ 
is the four-velocity of the particle.
From these two equations, one can solve out the locus of the orbiting particle
\ba
x_*^+ &=& {1-z_0\over z_0}\,{r_+E-r_-L/\ell\over r_+^2-r_-^2}\,\tau + x_0^+
\\[5pt]
x_*^- &=& (1-z_0)\,{r_+L/\ell-r_-E\over r_+^2-r_-^2}\,\tau + x_0^-
\ea
where $x_0^\pm$ are two constants and could be chosen to be vanishing
for convenience. The energy and the angular momentum is constrained by the 
mass-shell condition \cite{mtw_gravitation},
\ba
-1=\ell^2 \left( -{z_0\over 1-z_0}\dot x_*^+{}^2 
+ {1\over 1-z_0}\dot x_*^-{}^2 \right)
\ea
where, without loss of generality, the star has been chosen to have unit mass.
In the following, we will take the energy of the orbiting particle to be zero,
$E=0$. So 
\ba
-1 = {1-z_0\over (r_+^2-r_-^2)^2}(r_+^2-r_-^2/z_0)L^2
\ea
From which one can see that, to obtain real angular momentum, $L$, the
orbiting particle must be close enough to the black hole outer horizon, i.e.,
$z_0$ should satisfy
\ba  \label{nh-condition}
z_0<{r_-^2\over r_+^2}
\ea
which gives no constraint on $z_0$ for extremal black hole, $r_+=r_-$.

In the absence of the orbiting particle, a scalar field in the black hole background
is described by the free action, 
$S_\Phi=-{1\over 2}\int d^3x\sqrt{-g}\,\nabla_\mu\Phi\nabla^\mu\Phi$.
After adding the orbiting star, and let it be coupled to the 
scalar field with the following interaction,
\ba
S_{\rm int} = 4\pi\lambda\int d\tau \int d^3x\,\delta^3(x-x_*(\tau))\,\Phi(x)
\ea
So the orbiting star, as source, will excite the scalar mode which will 
propagate through the spacetime. The classical equation of motion of the 
scalar field reads
\ba
\sqrt{-g}\,\square\Phi =- 4\pi\lambda\int d\tau\delta^3(x-x_*(\tau))
\ea
The integral of the source term on the right hand side could be performed
explicitly,
\ba
-4\pi\lambda\,{(r_+^2-r_-^2)\ell\over (1-z_0)L}\,
\delta\left({r_+}x^++{1\over z_0}r_-x^-\right)\delta(z-z_0)
\ea
So this source respects a Killing symmetry, 
\ba  \label{circular-killing}
\chi = {1\over z_0}r_-\partial_+ - r_+\partial_-
\ea
To conform to this Killing symmetry, one can make the following 
variable separating assumption for the scalar field,
\ba
\Phi(x) &=& \sum_j e^{-ij(r_+x^+ +r_-x^-/z_0)}R_j(z)
\equiv \sum_j e^{-i(k_+x^+ + k_-x^-)}R_j(z)
\nonumber\\[5pt]
&=&\sum_j e^{-i(\omega t-m\phi)}R_j(z)
\ea
where $j\in\mathbb{Z}$, and we have defined $k_+=jr_+$, $k_-=jr_-/z_0$.
In above, through matching the phase, we have 
identified the the energy and angular momentum of the mode excited by the star
as $\omega=j{r_+^2 - r_-^2/z_0\over\ell^2}$ and
$m=j(1-{1\over z_0}){r_+r_-\over\ell}$. One notices that
both $\omega$ and $m$ depend on the position $z_0$ of the star.

Substituting into the equation
of motion, one find the following differential equation for the radial
function,
\ba  \label{eq-with-source}
z(1-z)R_j''(z) + (1-z)R'_j(z) +\left({k_+^2\over 4z} - {k_-^2\over 4}\right)R_j(z)
=-\tilde\lambda\,\delta(z-z_0)
\ea
where $\tilde\lambda={(r_+^2-r_-^2)\over (1-z_0)}{\lambda\over L}$.
This is an non-homogeneous differential equation. For non-integer $k_+$,
the two linearly independent solutions of the homogeneous equation, $\lambda=0$, are
given by the hyper-geometric functions,
\ba
R_j^{(1)}(z)&=&z^{-ik_+/2}\,{}_2F_1\left({k_++k_-\over 2i},{k_+-k_-\over 2i},1-ik_+,z\right)
\\[5pt]
R_j^{(2)}(z)&=&z^{ik_+/2}\,{}_2F_1\left(-{k_+-k_-\over 2i},-{k_++k_-\over 2i},1+ik_+,z\right)
\ea
Their asymptotic behaviors near the horizon are
\ba
R_j^{(1)}(z\to 0)\to z^{-ik_+/2} \,,\qquad
R_j^{(2)}(z\to 0)\to z^{ik_+/2}
\ea
Using the constraint condition (\ref{nh-condition}) on $z_0$, one can show
that $R_j^{(2)}(z)$ is purely ingoing near the horizon and 
$R_j^{(1)}(z)$ is purely outgoing.
Near the boundary,
\ba
R_j^{(1)}(z\to 1)&\to& {\Gamma(1-ik_+)\over\Gamma(1+{k_++k_-\over 2i})
\Gamma(1+{k_+-k_-\over 2i})} + {\cal O}(1-z)
\\[5pt]
R_j^{(2)}(z\to 1)&\to& {\Gamma(1+ik_+)\over\Gamma(1-{k_++k_-\over 2i})
\Gamma(1-{k_+-k_-\over 2i})} + {\cal O}(1-z)
\ea
which shows that the two homogeneous
solutions are not fall-off near the boundary. To obtain
a solution fall-off near the boundary, we can consider a
linear combination of the two,
\ba
R_j^{(\rm f)}(z) = c_j\, R_j^{(1)}(z) - c_j^*\, R_j^{(2)}(z)
\ea
where the complex number $c_j={\Gamma(1+{k_++k_-\over 2i})
\Gamma(1+{k_+-k_-\over 2i})/\Gamma(1-ik_+)}$. 
This `far' solution has the fall-off behavior
near the boundary, which is called Newmann boundary condition in \cite{ps_kerr_cft},
\ba
R_j^{(\rm f)}(z\to 1) = i\pi{k_+^2-k_-^2\over 4}
{\sinh(\pi k_+)\over\sinh({k_++k_-\over 2}\pi)\sinh({k_+-k_-\over 2}\pi)}(1-z)
+ {\cal O}((1-z)^2)
\ea
Since we want to calculate the decay rate, so only solution which is purely
ingoing near the horizon is desired. In all, the solution of the 
inhomogeneous equation (\ref{eq-with-source}) with proper asymptotic behaviors 
can be constructed as follows
\ba
R_j(z) = \theta(z_0-z)\, c_{\rm n}\, R_j^{(\rm n)}(z) 
+ \theta(z-z_0)\, c_{\rm f}\, R_j^{(\rm f)}(z) 
\ea
where $R_j^{(\rm n)}(z) =R_j^{(2)}(z) $, $c_{\rm n}$ and $c_{\rm f}$ are
two constants to be determined, $\theta(x)$ is the usual step function.
Substituting into the equation of motion, the coefficients can be determined,
\ba
c_{\rm n} = {\tilde\lambda\over 2}\,
{R_j^{(\rm f)}(z_0)\over W}\,,\qquad
c_{\rm f} = {\tilde\lambda\over 2}\,
{R_j^{(\rm n)}(z_0)\over W}\,,\qquad
\ea
where $W$ is $z$-independent Wronskian,
\ba
W = z\big(R_j^{(\rm f)}(z)R_j^{(\rm n)}{}'(z)-R_j^{(\rm f)}{}'(z)R_j^{(\rm n)}(z)\big)
= {\Gamma(1+{k_++k_-\over 2i})
\Gamma(1+{k_+-k_-\over 2i})\over\Gamma(-ik_+)}
\ea
With the above solution, one can calculate the Klein-Gordon particle
number flux 
\ba
{\cal F} = - \int d\phi \sqrt{-g} J^r
=- \int d\phi\, r {\partial r\over \partial z} J^z d\phi
\ea
with the current
\ba
J_\mu={i\over 8\pi}(\Phi^*\nabla_\mu\Phi-\Phi\nabla_\mu\Phi^*)
\ea
The flux of mode-$j$ is then 
\ba
{\cal F}_j = -i{r_+^2-r_-^2\over 2\ell^2} 
z\big(R_j^*{}'(z)R_j(z) - R_j^*(z)R_j'(z)\big)
\ea
Near the boundary, the flux vanishes, ${\cal F}_j(z\to 1) = 0$.
Near the horizon, 
\ba
{\cal F}_j(z\to 0) = {(r_+^2-r_-^2)k_+\over 2\ell^2}\, |c_{\rm n}|^2
\ea
So the decay rate of the particle, with energy $\omega$ and angular
momentum $m$, excited by the source is  \cite{ms_low_energy_dynamics}
\ba
{\cal R}_j^{\rm (btz)} = {1\over e^{\omega-m\Omega_{\rm H}\over T_{\rm H}}-1}
{\cal F}_j(z\to 0)
\ea
where $\Omega_{\rm H}$ is the angular velocity of the horizon, \cite{poisson_book}
\ba
\Omega_{\rm H}= - {g_{t\phi}\over g_{\phi\phi}}\Bigg|_{r=r_+}
={r_-\over \ell\, r_+}
\ea
Simple calculation shows that 
$\omega-m\Omega_{\rm H}=j{r_+^2-r_-^2\over\ell^2}=2\pi j r_+ T_{\rm H}$,
so another form of the decay rate is
\ba  \label{btz-decay-rate}
{\cal R}_j^{\rm (btz)}={1\over e^{2\pi k_+}-1}{\cal F}_j(z\to 0)
\ea
An interesting fact to notice is that, the factor of the occupation number
is independent of the position $z_0$ of the star, although
the energy $\omega$ and angular momentum $m$ of the excitation
depend on $z_0$ separately. One can find that this is important for
matching of decay rates in BTZ/CFT correspondence as shown in the following.

\subsection{Decay rates on the CFT side}

The pure BTZ black hole dynamics will correspond to a CFT defined on the boundary
which is controlled by two copies of Virasoro algebras corresponding
to left- and right-mover respectively. If the CFT is described by an
action, $S_{\rm CFT}$, then when we add to the BTZ black hole a particle
which is circulating the black hole, the action describing the CFT will be changed
to, $S_{\rm CFT}+S_{\rm CFT}^{\rm int}$, 
\ba
S_{\rm CFT}^{\rm int} = \int d\sigma^+ d\sigma^-\, 
J(\sigma^+,\sigma^-)O(\sigma^+,\sigma^-)
\ea
where $\sigma^\pm=t/\ell\pm\phi$, and 
$t, \phi$ are just the Schwarzschild coordinates of BTZ black hole space time.
Since we consider only the coupling of the scalar field to the source 
(orbiting particle), so one expect the operator $O$ is the CFT dual to the
scalar field, and the source $J$ is induced from the orbiting star.

To determine the source $J$, one should at first note that it should
respect the Killing symmetry (\ref{circular-killing}) which is respected 
by the orbiting particle. That is we must have the following mode expansion
\ba
J(\sigma^+,\sigma^-)&=&\sum_j e^{-ij(r_+x^+ +r_-x^-/z_0)}J_j
= \sum_j e^{-i(k_+x^+ + k_-x^-)}J_j
\ea
In terms of $\sigma^\pm$,
\ba
x^\pm={r_+-r_-\over 2\ell}\sigma^+\pm {r_++r_-\over2\ell}\sigma^-
\ea
the source can then be written as,
\ba
J(\sigma^+,\sigma^-) = \sum_j e^{-i(p_+\sigma^+ + p_-\sigma^-)}J_j
\ea
where $p_\pm\equiv{(k_+\pm k_-)(r_+\mp r_-)\over 2\ell}$ is defined.
The interaction action then reads
\ba
S_{\rm CFT}^{\rm int} 
=\sum_j J_j \int d\sigma^+ d\sigma^-\, e^{-i(p_+\sigma^+ + p_-\sigma^-)}
O(\sigma^+,\sigma^-)
\ea
The vacuum-vacuum transition rate excited by the source is
\ba
{\cal R} = 2\pi \sum_j |J_j|^2 \int d\sigma^+ d\sigma^-\, 
e^{-i(p_+\sigma^+ + p_-\sigma^-)}\langle O^\dag(\sigma^+,\sigma^-)O(0,0)\rangle
\ea
The angle-bracket is two-point correlation function of operator $O$ in
the unperturbed CFT. The two point function of ${\cal O}$ and 
its Fourier transform can be found in references, e.g., 
\cite{bhss_superradiance_kerr_cft}. For the scalar case, the result is 
\ba \label{fourier-oo}
&&\int d\sigma^+ d\sigma^-\, 
e^{-i(p_+\sigma^+ + p_-\sigma^-)}\langle O^\dag(\sigma^+,\sigma^-)O(0,0)\rangle
\nonumber\\[5pt]
&=&C_O^2(2\pi T_R)(2\pi T_L) e^{-{p_-\over 2T_R}-{p_+\over 2T_L}}
\left|\Gamma\left(1+i{p_-\over 2\pi T_R}\right)
\Gamma\left(1+i{p_+\over 2\pi T_L}\right)\right|^2
\ea
where the overall normalization constant $C_O$ is not determined merely by
the conformal invariance, and the left- and right-temperatures are defined by
\cite{bss_qnm}
\ba  \label{non-chiral-cft-temp}
T_L={(r_+-r_-)\over 2\pi\ell}\,,\qquad T_R={(r_++r_-)\over 2\pi\ell}\,,
\ea
Now the transition rate reads
\ba
{\cal R}^{\rm (cft)} = 2\pi(r_+^2-r_-^2) \sum_j C_O^2|J_j|^2\, e^{-\pi k_+}
\left|\Gamma\left(1-{k_++k_-\over 2i}\right)
\Gamma\left(1-{k_+-k_-\over 2i}\right)\right|^2
\ea
The next task is to determine the source $J_j$ which is related to the 
orbiting star in the bulk.

To do that, one extends the near horizon solution 
$R_j(z<z_0)$ to the whole spacetime in the bulk, 
$R_j^{\rm ext}(z)=c_{\rm n}R_j^{(\rm n)}(z)$, ($0<z<1$), 
and extract $J_j$ from its Dirichlet mode on the boundary 
(that is, the term fall-off as $(1-z)^0$ near boundary) \cite{ps_kerr_cft}
\ba
R_j^{\rm ext}(z\to 1)
=c_{\rm n}{\Gamma(1+ik_+)\over\Gamma(1-{k_++k_-\over 2i})
\Gamma(1-{k_+-k_-\over 2i})}
\equiv J_j
\ea
With this source, the transition rate of mode-$j$ can be written as
\ba
{\cal R}_j^{\rm (cft)} &=& 2\pi(r_+^2-r_-^2)|c_{\rm n}|^2e^{-\pi k_+}
\left|\Gamma(1+ik_+)\right|^2C_O^2
\nonumber\\[5pt]
&=&2\pi^2(r_+^2-r_-^2)k_+|c_{\rm n}|^2 {e^{-\pi k_+}\over\sinh(\pi k_+)}C_O^2
\nonumber\\[5pt]
&=& 8\pi^2\ell^2C_O^2 {1\over e^{2\pi k_+}-1}{\cal F}_j(z\to 0)
\ea
which equals precisely to the result computed in the gravity side
(\ref{btz-decay-rate}) if the coefficient is chosen to be $C_O^2=1/(8\pi^2\ell^2)$.

\section{Near-horizon-extremal BTZ/Chiral CFT}

In this section, we will compute the decay rate of scalar wave 
in the extremal BTZ spacetime and compare it to that computed in the CFT.
The correspondence is Near-horizon-extremal BTZ/Chiral 2D-CFT (NHEBTZ/$\chi$CFT)
\cite{strominger_ads2_qg,bbjs_XCFT_Ebh}. Before doing the computation, 
let us at first review some facts about this correspondence.

The metric of the extremal BTZ black hole can be obtained by setting
$r_+=r_-=r_{\rm H}$ in that of non-extremal one (\ref{btz-metric}):
\beq
ds^2 = -{(r^2-r_{\rm H}^2)^2\over r^2\ell^2}dt^2 
+ {r^2\ell^2\over (r^2-r_{\rm H}^2)^2}dr^2
  + r^2\left( d\phi -{r_{\rm H}^2\over r^2\ell} dt\right)^2
\label{ext-btz-metric}
\eeq
For the extremal BTZ black hole, its Hawking temperature is zero, $T_{\rm H}=0$.
Define null coordinates, $\hat u=t/\ell-\phi$, $\hat v=t/\ell+\phi$,
and a new radial coordinate $r^2-r_{\rm H}^2=\ell^2\hat y^2$. In terms
of the these new coordinates, the metric reads
\ba
ds^2={\ell^2\over 4}{{d\hat y}^2\over{\hat y}^2}
-\ell^2 {\hat y}\,d{\hat u}d{\hat v}+r_{\rm H}^2d{\hat u}^2
\ea
One should note that the original periodic condition of the angular
variable $\phi$ will leads to the following periodic conditions of the
null variables,
\ba
(\hat u, \hat v) \approx (\hat u-2\pi, \hat v+2\pi)
\ea
It is pointed out that it is the quantum gravity on the near horizon
region of the extremal BTZ spacetime that will correspond to a (chiral)
two-dimension CFT on the boundary \cite{bbjs_XCFT_Ebh}. The near
horizon limit can be obtained by defining
\ba  \label{nhebtz-coord}
\hat u={\ell\over r_{\rm H}} u\,,\quad 
\hat v={1\over \epsilon}{r_{\rm H}\over\ell}v\,,\quad
\hat y=\epsilon y
\ea
and letting $\epsilon\to 0$, and keeping $u, v, y$ fixed. The important 
point is that on this near-horizon portion of the extremal BTZ spacetime,
the periodic conditions of the null variables, $u, v$, become chiral:
\ba
(u, v) \approx (u-2\pi{r_{\rm H}\over \ell}, v)
\ea
That is, only the $u$ has finite period, and $\partial_u$ will be matched
to the left-translation on the (chiral) CFT and related to the finite left
temperature,
\ba
T_L = {r_{\rm H}\over \pi \ell}
\ea
which can be obtained apparently by letting $r_+=r_-=r_{\rm H}$
in non-extremal case discussed in the previous section (\ref{non-chiral-cft-temp}).
$\partial_v$ will be matched to the right-translation on the (chiral) CFT and 
related to the vanishing right temperature, $T_R=0$. So one can make the following
identification
\ba
u={r_{\rm H}\over\ell}\sigma^+\,,\quad
v=\sigma^-
\ea
where $\sigma^\pm=\tau\pm\sigma$ are coordinates of the CFT defined on
a (chiral) strip with normalized identification, 
$(\sigma^+,\sigma^-)\approx(\sigma^++2\pi,\sigma^-)$ \cite{cardy_occf}. 

In terms of the near horizon coordinate (\ref{nhebtz-coord}), the black hole
metric reads
\ba
ds^2&=&{\ell^2\over 4}{{dy}^2\over{y}^2}
-\ell^2 {y}\,d{u}d{v}+\ell^2d{u}^2
\nonumber\\[5pt]
&=&{\ell^2\over 4}\left(-y^2dv^2+{dy^2\over y^2}\right)
+\ell^2\left(du-{y\over 2}dv\right)^2
\label{nhebtz-met}
\ea
which is an $S^1$-fibration on $AdS_2$. The fibre direction $u$, 
similar to the near-horizon-extremal RN black hole/CFT correspondence 
\cite{hmns_cft_Ebh}, is
mapped to the left-direction $\sigma^+$ of the chiral 
CFT as stated in the previous paragraph. The near-horizon geometry (\ref{nhebtz-met})
has isometry group,
\ba
SL(2,\mathbb{R})\times U(1)\,,
\ea
where the factor $SL(2,\mathbb{R})$ is generated by the following Killing 
vectors \cite{strominger_ads2_qg},
\ba  \label{sl2r-isometry}
H_{-1}=\partial_v\,,\quad
H_0=v\partial_v-y\partial_y\,,\quad
H_{+1}=\left(v^2+{1\over y^2}\right)\partial_v-2vy\partial_y+{1\over y}\partial_u
\ea
and the $U(1)$ factor is generated by
\ba
Q_0=\partial_u
\ea
As pointed out in \cite{bbjs_XCFT_Ebh}, in the NHEBTZ/$\chi$CFT 
correspondence, the $SL(2,\mathbb{R})$ isometries are realized trivially on the
boundary CFT: it is associated to the re-parametrization of the non-compact $v$
on the boundary. The physical states do not carry global charges of 
$SL(2,\mathbb{R})$, they only carry the $u$-momentum, 
$Q_0$, which is the zero mode of a full Virasoro algebra. This `chiral' correspondence
   \footnote{
     I.e., only one Virasoro symmetry acting on the boundary CFT.
   }
is very different 
from other Gravity/CFT correspondences, e.g., the usual AdS$_3$/CFT$_2$
correspondence \cite{bh_ads3_cft2} and Kerr/CFT correspondence
(see Ref. {[18]} and references therein);
the asymptotic symmetry group (ASG) includes
two Virasoro algebras. For example, in
Kerr/CFT, the isometry group of the near horizon geometry is
$SL(2,\mathbb{R})\times U(1)$, however, in the ASG,
the zero-mode of the L Virasoro algebra 
corresponds to the $U(1)$ isometry group \cite{ghss_kerr_cft};
for the R Virasoro algebra, there is an $SL(2,\mathbb{R})$ subgroup that can be
identified as the isometry group \cite{cl_kerr_cft}. (For the discussion of the symplectic symmetry
algebra, which is different from the ASG, please refer to Ref. \cite{chss_ssg}.)

\subsection{Decay rate on the gravity side}

Just like the non-extremal black hole case discussed in the previous section, 
we will compute the decay rate of the scalar wave excited by an orbiting star at the
near horizon region of the extremal BTZ spacetime. In the BTZ spacetime
two Killing vectors $\partial_t$ and $\partial_\phi$ become
$\xi_{(u)}=\partial_u$ and $\xi_{(v)}=\partial_v$ in the near-horizon
region (\ref{nhebtz-met}), there are two corresponding conserved quantities,
\ba
\ell^2C_u &=& -g_{\mu\nu}\dot x_*^\mu\xi^\nu_{(u)}
={\ell^2\over 2}y_0 \dot u_*
\\[5pt]
\ell^2C_v &=& g_{\mu\nu}\dot x_*^\mu\xi^\nu_{(v)}
=\ell^2 \dot u_* - {\ell^2\over 2}y_0 \dot v_*
\ea
where $u_*, v_*$ are orbiting-coordinates of the star,
the constant $y_0$ is the radial position of the star. 
The solution is given by
\ba
u_* = {2C_u\over y_0} \tau + u_0\,,\qquad
v_* = {2\over y_0}\left({2C_u\over y_0}-C_v\right)\tau + v_0
\ea
with $\tau$ the proper time. Two constants $C_u$ and $C_v$ are 
linear combination of the `energy' $E$ and `angular momentum' $L$ of the star,
they are further constrained by the on-shell condition of the star (e.g., with unit
mass without loss of generality),
$-1=-\ell^2 {y_0}\,\dot u_*\dot v_*+\ell^2\dot u_*^2$. 

Let the star be coupled to the massless scalar field, the corresponding
action reads,
\ba
S=S_{\rm EH}-{1\over 2}\int d^3x\,\sqrt{-g}\nabla_\mu\Phi\nabla^\mu\Phi
+\lambda\int d\tau\int d^3x\,\delta^3(x-x_*)\Phi(x)
\ea
$S_{\rm EH}$ is the Einstein-Hilbert action. The equation of motion for the scalar
field is
   \footnote{
     One simple calculation shows that
     $$\sqrt{-g}\,\square\Phi =
     {\ell^3\over 4}\left[H_0^2-{1\over 2}(H_{-1}H_{+1}+H_{+1}H_{-1})\right]\Phi$$
     where the operator on the right-hand side is just the 2nd-order Casimir operator 
     of $SL(2,\mathbb{R})$ up to a constant factor. Now according to a similar
     analysis in \cite{strominger_ads2_qg}, a general scalar field solution
     of homogeneous equation with $\lambda=0$, 
     $\sqrt{-g}\,\square\Phi=0$, can be expanded 
     in terms of the highest weight solution $\phi_h$,
     which satisfies $H_{+1}\phi_h=0$ and $H_0\phi_h=h\phi_h$, and
     their descendants $D_n=(H_{-1})^n\phi_h$ (with $n=1,2,\dots$), 
     $\Phi=\sum_{n=0} a_nD_n+{\rm h.c.}$. 
     This comes from that fact that $H_{-1}$ commutes with 
     the 2nd-order Casimir operator. Furthermore, corresponding to
     each highest weight solution, there is a primary operator ${\cal O}_h$
     in the boundary field theory. The non-homogeneous solution
     for $\lambda\neq 0$ will be constructed from the homogeneous
     solutions via the Green function methods which will be discussed in 
     the following.
   }
\ba
\sqrt{-g}\,\square\Phi
=-{y_0^2\over 4 C_u}\lambda\,\delta(u-{y_0\over 2}v)\delta(y-y_0)
\ea
The source has Killing symmetry, 
$\chi=\partial_u+{2\over y_0}\partial_v=Q_0+{2\over y_0}H_{-1}$, 
which is the linear combination of the generator of $U(1)$ and
one generator of $SL(2,\mathbb{R})$.
Since there is an expansion of Dirac delta function as sum,
$\delta(x)={1\over 2\pi}\sum_{n\in\mathbb{Z}}e^{inx}$,
we can separate the variables of the scalar field by
\ba
\Phi =\sum_j e^{ij(u-{y_0\over 2}v)}R_j(y) \equiv
\sum_j e^{iju-i\omega v}R_j(y)
\ea
where $j= m{\ell\over r_{\rm H}}$ with $m\in\mathbb{Z}$, 
$\omega=j{y_0\over 2}=m{y_0\ell\over 2r_{\rm H}}$. Substituting this expansion into
the equation of motion, one find the radial equation,
\ba
(y^2 R_j'(y))' + \left({\omega^2\over y^2}-{j\omega\over y}\right)R_j(y) 
= -\tilde\lambda\delta(y-y_0)
\ea
where $\tilde\lambda={y_0^2\,r\big._{\!\!\rm H}\over 2\pi\ell^4C_u}\lambda$. Two
linearly independent solutions of the homogeneous equation, $\lambda=0$, are
given by the hypergeometric functions:
\ba
R_j^{(1)}(y) 
&=& {e^{i\omega/y} \over y}\,{}_1F_1\left(1+{ij\over 2},2,-{2i\omega\over y}\right)
\\[5pt]
R_j^{(2)}(y) 
&=& {e^{i\omega/y} \over y}\,U\left(1+{ij\over 2},2,-{2i\omega\over y}\right)
\ea
The asymptotic behavior near the horizon, $y\to 0$,
\ba
R_j^{(1)}(y) 
&\sim& e^{i\omega/y} y^{ij/2}{(2i\omega)^{-1-ij/2}\over\Gamma(1-ij/2)}
+ e^{-i\omega/y} y^{-ij/2}{(-2i\omega)^{-1+ij/2}\over\Gamma(1+ij/2)}
\\[5pt]
R_j^{(2)}(y) 
&\sim& e^{i\omega/y} y^{ij/2}(-2i\omega)^{-1-ij/2}
\ea
which shows that $R_j^{(2)}$ is purely ingoing near the horizon. Near the boundary
of the near horizon region, $y\to\infty$,
\ba  \label{asy-R1}
R_j^{(1)}(y) 
&\sim& y^{-1}
\\[5pt]\label{asy-R2}
R_j^{(2)}(y) 
&\sim& {1\over j\omega\Gamma({ij\over 2})}
+ (A+B\log y)y^{-1}
\ea
where $A={i-j+2j\gamma_E+j\log(-2i\omega)+j\psi^{(0)}(1+ij/2)\over j\Gamma(ij/2)}$, $B=-{1\over \Gamma(ij/2)}$ ($\gamma_E$ is the Euler constant,
$\psi^{(0)}(x)=\Gamma'[x]/\Gamma[x]$). The above asymptotic behavior shows
that $R_j^{(1)}$ satisfies pure Neumann boundary condition.

In order to describe ingoing particle flux of black hole, 
one can construct the proper solution of the (in-homogeneous) equation of motion
as Ref.\cite{ps_kerr_cft},
\ba
R_j(y) = \theta(y_0-y)C_2 R^{(2)}_j(y) + \theta(y-y_0)C_1 R^{(1)}_j(y)
\ea
which is purely ingoing near the horizon and satisfies pure Neumann boundary
condition near the boundary. The constants $C_{1,2}$ are
\ba
C_1={\tilde\lambda\over 2}{R_j^{(2)}(y_0)\over W}\,,\qquad
C_2={\tilde\lambda\over 2}{R_j^{(1)}(y_0)\over W}
\ea
where the Wronskian is defined by,
\ba
W=y^2(R_j^{(1)}(y)R_j^{(2)}{}'(y)-R_j^{(2)}(y)R_j^{(1)}{}'(y))
={1\over j\omega\Gamma({ij\over 2})}
\ea
which is $y$-independent constant.

Now we can compute the Klein-Gordon particle number flux per unit time,
\ba
{\cal F} = -\int\sqrt{-g}\, J^y du
\ea
where the current $J_\mu$ is defined by, $J_\mu={i\over 8\pi}(\Phi^*\nabla_\mu\Phi-h.c.)=-{1\over 4\pi}{\rm Im}(\Phi^*\nabla_\mu\Phi)$.
The flux of mode-$j$ is
\ba
{\cal F}_j ={r_{\rm H}\over 2}y^2{\rm Im}(R_j^*\partial_yR_j)
\ea
Near boundary, ${\cal F}_j(y\to\infty)=0$. Near the horizon, 
\ba \label{flux-gr-side}
{\cal F}_j(y\to 0) = {r_{\rm H}\over 8|\omega|}e^{\pi j/2}|C_2|^2
\ea
for $\omega=jy_0/2<0$.

\subsection{Decay rate on the CFT side}

The introduction of the star into the gravity system will
lead to the perturbation of the CFT action which is expected by
the Gravity/CFT correspondence. Similar to the non-extremal case
discussed in the previous section, the unperturbed CFT action,
$S_{\rm CFT}$, is changed to $S_{\rm CFT}+S_{\rm CFT}^{\rm int}$,
where
\ba
S_{\rm CFT}^{\rm int} 
= \int d\sigma^+ d\sigma^- J(\sigma^+,\sigma^-){\cal O}(\sigma^+,\sigma^-)
\ea
The source $J$ will be determined by the asymptotic behavior of the scalar field,
so it could be expanded just like the scalar field,
\ba
J(\sigma^+,\sigma^-)=\sum_j e^{iju-i\omega v} J_j
=\sum_j e^{i(m\sigma^+-\omega\sigma^-)}J_j
\ea
where $m={jr_{\rm H}\over\ell}\in\mathbb{Z}$ as defined in 
the previous subsection, 
and we have used the NHEBTZ/$\chi$CFT dictionary discussed 
in the beginning of this section, $u={r_{\rm H}\over\ell}\sigma^+$, $v=\sigma^-$.
Furthermore, from the asymptotic behavior of the scalar fields,
(\ref{asy-R1}) and (\ref{asy-R2}), the conformal weight of the operator
${\cal O}$ can be read off, $h_L=h_R=1$.

The decay rate of vacuum-to-vacuum per unit time is
\ba
{\cal R} &=& 2\pi\sum_j|J_j|^2\int d\sigma^+ d\sigma^-
e^{i(m\sigma^+-\omega\sigma^-)}\langle{\cal O}(\sigma^+,\sigma^-)
{\cal O}(0,0)\rangle
\nonumber\\[5pt]
&=& 2\pi\sum_j|J_j|^2C_O^2(2\pi T_R)(2\pi T_L) 
e^{-{\omega\over 2T_R}+{m\over 2T_L}}
\left|\Gamma\left(1+i{\omega\over 2\pi T_R}\right)
\Gamma\left(1-i{m\over 2\pi T_L}\right)\right|^2
\nonumber\\
\ea
where we have used (\ref{fourier-oo}) and $C_O$ is a normalization constant.
Using $m=jr_{\rm H}/\ell$, $T_L=r_{\rm H}/(\pi\ell)$ and taking the limit
$T_R\to 0$, the decay rate becomes,
\ba
{\cal R} &=& 2\pi\sum_j|J_j|^2C_O^2{2r_{\rm H}\over\ell}
e^{\pi j/2}|\Gamma(1-ij/2)|^2\,2\pi|\omega|
\ea
for $\omega<0$.

To determine the source $J_j$, as done in the non-extremal case, we
can extend the near horizon solution to the whole
spacetime region, 
$R_j(y<y_0)\to R_j^{\rm ext}(y)=C_2 R_j^{(2)}(y), (0<y<\infty)$. 
The asymptotic behavior of this extended solution reads, see (\ref{asy-R2}),
\ba
R_j^{\rm ext}(y\to\infty)\sim C_2 {1\over j\omega\Gamma({ij\over 2})}
+ C_2(A+B\log y)y^{-1}
\ea
Then $J_j$ is the coefficient of $y^0$ (the boundary Dirichlet mode),
\ba
J_j = {C_2\over j\omega\Gamma({ij\over 2})}
\ea 
With this source, the decay rate is
\ba  \label{dr-cft-side}
{\cal R} = C_O^2{2\pi^2 r_{\rm H}\over|\omega|\ell}e^{\pi j/2}|C_2|^2
\ea
for $\omega<0$. This decay rate is precisely matched with 
(\ref{flux-gr-side}) computed in gravity side if we take
$C_O^2={\ell\over 16\pi^2}$.

\section{Final remarks}

We have shown that the correspondence between the BTZ black hole and the
CFT defined on the boundary ensures that the decay rates of massless scalar waves
computed on both sides match precisely. Two cases are discussed.
The first case is for non-extremal BTZ/non-chiral CFT correspondence. 
A point should be mentioned is that since the temperature is finite
for non-extremal BTZ black hole, so there is a Planck factor (occupation number) 
appearing in the expression of the decay rate. 
The second case is for the near-horizon-extremal BTZ/chiral 2D-CFT correspondence.
Since the temperature is zero in this case, there is no
Planck factor in the expression of the decay rate. 
This is similar to the cases discussed in \cite{ps_kerr_cft,bgx_scalar_KN}:
the Planck factor is trivial in the extremal black hole spacetime.

As shown in \cite{ps_kerr_cft, bgx_scalar_KN}, it is interesting to examine whether
other types of excitations give further evidences to the BTZ/CFT correspondence:
gravity waves, vector waves, etc.. These questions will be explored in other places.

\bigskip\bigskip

\noindent
{\bf Acknowledgments.}
R. L. and J.-K. Z. are supported by National Natural Science Foundation of China
(Grant No. 11205048). 
R. L. is also supported by the Foundation for Young Key Teacher of 
Henan Normal University.

\end{document}